\documentclass[aps,prb,reprint,twocolumn,longbibliography,superscriptaddress,draft=false]{revtex4-2}
\usepackage{hyperref}
\hypersetup{colorlinks=true,
		    linkcolor=blue,
		    citecolor=blue,
		    allcolors=blue}
\usepackage{natbib}
\usepackage{amsmath, amsfonts, bm, bbm, braket, mathtools}
\usepackage{array}
\usepackage{graphicx,overpic}
\graphicspath{{./}}
\usepackage{pifont}% http://ctan.org/pkg/pifont
\newcommand{\mytexttilde}{\raise.17ex\hbox{$\scriptstyle\sim$}}

\begin{document}

%\preprint{}

%\title{Role of  interlayer spin interactions in $\alpha$-RuCl$_3$ }
\title{Intermediate phases in $\alpha$-RuCl$_3$ under in-plane magnetic field via interlayer spin interactions}
\author{Jiefu Cen}
\affiliation{Department of Physics, University of Toronto, Toronto, Ontario, Canada M5S 1A7}
\author{Hae-Young Kee}
\email[]{hykee@physics.utoronto.ca}
\affiliation{Department of Physics, University of Toronto, Toronto, Ontario, Canada M5S 1A7}
\affiliation{Canadian Institute for Advanced Research, CIFAR Program in Quantum Materials, Toronto, Ontario, Canada, M5G 1M1}
\begin{abstract}
$\alpha$-RuCl$_3$ has attracted significant attention as a prime candidate for the spin-1/2 Kitaev spin liquid in two-dimensional honeycomb lattices. Although its ground state is magnetically ordered, the order is suppressed under a moderate in-plane magnetic field. The intermediate regime of the field has exotic behaviors, some of which are claimed to originate from a Kitaev spin liquid. In resolving debates surrounding these behaviors, interlayer interactions in $\alpha$-RuCl$_3$ have been largely overlooked due to their perceived weakness in van der Waals materials. However, near the transition, they may become significant as the field energy approaches the interlayer coupling scale. Here we investigate the effects of interlayer couplings in $\alpha$-RuCl$_3$ with $R\bar{3}$ and $C2/m$ structures. We first examine their effects on the transition temperature ($T_N$) using classical Monte Carlo simulations. We found that the interlayer couplings have minimal effects on $T_N$, and the different $T_N$ between the two structures are mainly due to the anisotropy in intralayer interactions.
Focusing on the $R{\bar 3}$ structure, we show that the nearest neighbor interlayer interaction is the XXZ type due to the symmetry, and the next nearest neighbor interaction of the Kitaev type is crucial for the transition between two zigzag orders under an in-plane field. 
%study the role of interlayer coupling in shaping the phase diagrams under in-plane magnetic fields. The nearest neighbor interlayer interaction is XXZ type due to the symmetry, and the next nearest neighbor of Kitaev interaction is crucial for the transition between two zig-zag orders under in-plane field. 
Furthermore, an intermediate phase with a large unit cell emerges due to the interlayer interactions. Our findings provide insights into the exotic behaviors and sample dependence reported in $\alpha$-RuCl$_3$.
\end{abstract}
%\date{\today}
\maketitle

\section{Introduction}
Kitaev quantum spin liquid, the ground state of the exactly solvable Kitaev model \cite{Kitaev2006}, has attracted much interest since the microscopic mechanism to generate the Kitaev interaction was uncovered in materials with large spin-orbit coupling (SOC) \cite{Jackeli2009PRL,Rau2014,Rau2016ARCMP,Winter2017,Motome2019JPSJ,Takagi2019NRP,Takayama2021JPSJ,Rouso2024RoPP}. A$_2$IrO$_3$ (A = Li, Na) with honeycomb lattice was proposed for candidate materials due to its strong SOC \cite{Jackeli2009PRL}. 
Later, $\alpha$-RuCl$_3$ emerged as a prime candidate material due to its quasi-two-dimensional (2D) structure and relatively strong SOC compared to the bandwidth \cite{Plumb2014PRB,HSKim2015PRB,Banerjee2016,matsuda2025arXiv}. Although the ground state of $\alpha$-RuCl$_3$ has the zigzag (ZZ) antiferromagnetic order \cite{Sears2015PRB,Johnson2015,HSKim2015PRB}, it can be suppressed under a moderate in-plane magnetic field (~7 T) \cite{Majumder2015,Kubota2015,Leahy2017,baek_evidence_2017,Sears2017,Wolter2017,Zheng2017PRL,kasahara2018thermal}.
In this regime, exotic phenomena have been reported, motivating claims of the Kitaev spin liquid. Notable examples include the seemingly half-quantized thermal Hall conductivity \cite{kasahara2018thermal} and the oscillatory field dependence observed in longitudinal thermal conductivity \cite{Czajka2021}. However, the reproducibility and origins of these results are still under debate \cite{Bruin2021,Yokoi2021,Lefran2022PRX,Kee2023NM,Yamashita2020PRB,Czajka2023NM,Lefran2023PRB,Bruin2022APL,Kasahara2022PRB,Suetsugu2022,Zhang2024PRL,matsuda2025arXiv}.

It is challenging to resolve these debates and the spin model for $\alpha$-RuCl$_3$ due to sample dependence and the complexity of the extended Kitaev model, which includes the Heisenberg interaction and another bond-dependent interaction known as the $\Gamma$ interaction \cite{Rau2014,HSKim2015PRB,HSKim2016PRB,Winter2016PRB,Rouso2024RoPP}.
Even in the face of challenges, progress continues. There is a consensus that $\alpha$-RuCl$_3$ undergoes a structural transition from the monoclinic $C2/m$ structure to the rhombohedral $R\bar{3}$ around 150K \cite{Kubota2015,widmann2019PRB,namba2024PRM,RuCl3_growth_Kim2022,RuCl3_R3_kim2023}.
High-quality samples have a single magnetic transition around $T_{N}=7$ K to the ZZ phase, while samples with many stacking faults have additional transitions from 10 K to 14 K \cite{Banerjee2016,Kubota2015,RuCl3_R3_kim2023,RuCl3_growth_Kim2022}. The effects of interlayer couplings have been observed and studied in the zero-field zigzag phase and the paramagnetic high-field phase \cite{BalzPRB2019,Lukas2020PRB}.
Furthermore, in the intermediate field regime before reaching the proposed Kitaev spin liquid phase, the ZZ order across layers undergoes a transition, changing its pattern from a three-unit-cell to a six-unit-cell configuration, denoted as ZZ$_1$ and ZZ$_2$, respectively \cite{balz_field_2021}, signaling the importance of the interlayer coupling under the magnetic field.

Despite hints of interlayer couplings' role under a magnetic field, interlayer spin interactions have been largely overlooked in spin models of $\alpha$-RuCl$_3$. In two-dimensional van der Waals (vdW) materials, these interactions are typically neglected due to their relatively weak strength, which has minimal impact on the ordered states stabilized by stronger intralayer anisotropic spin interactions \cite{HSKim2016PRB}.
However, in regions near the transition to partially polarized states, where the energy scale of the field is comparable to the interlayer interaction and multiple competing phases arise due to frustrated Kitaev and $\Gamma$ interactions, interlayer interactions may play an important role and provide a natural explanation for the sample-dependent behavior of $\alpha$-RuCl$_3$. 
These dependencies include variations in crystal structure, magnetic critical temperature, and thermal conductivity, as they are sensitive to modifications in the vdW layer structure.

In this paper, we study the role of interlayer interactions in $\alpha$-RuCl$_3$ in determining the critical temperatures and magnetic field-driven phase transitions. In particular, we focus on whether interlayer interactions can induce an intermediate phase (IP) between the ZZ and polarized states. To explore their effects, we first derive a minimal interlayer exchange interaction model based on the symmetries of the $R\bar{3}$ and $C2/m$ structure with interaction strengths guided by
\textit{ab initio} calculations together with the strong coupling expansion theory. The transition temperatures in the $R\bar{3}$ and $C2/m$ structures and the phase diagrams under the in-plane magnetic fields are then studied using classical Monte Carlo (CMC) simulations.

Our main findings are as follows. First, small interlayer interactions have a small impact on $T_N$ in the zero field, as one may expect, since the zigzag order is stabilized by the strong 2D intralayer interactions. The $C2/m$ structure has a much higher $T_N$ than the $R\bar{3}$ structure because of the bond anisotropy of the intralayer interactions, while the interlayer interactions have minimal effects. 
Second, the observed 3D magnetic transition in the $R{\bar 3}$ structure under an in-plane magnetic field along the $\hat{a}$-axis (perpendicular to one of the bonds), namely from ZZ$_1$ to ZZ$_2$ in Ref. \cite{balz_field_2021}, is due to the competition between the XXZ-type nearest neighbor (NN) and Kitaev-type second NN interlayer interactions. 
Third, IPs emerge between the ZZ$_2$ and polarized phases as a result of interlayer interactions. Because these phases are sensitive to layer stacking, this provides a natural explanation for the sample dependence observed in the magnetic anomalies \cite{Yamashita2020PRB, Zhang2023PRM,Zhang2024PRM} and potentially for the non-monotonic behavior seen in longitudinal thermal conductivity \cite{Bruin2022APL, Lefran2023PRB}.

The paper is organized as follows. In Sec. II, we derive interlayer spin interactions for the $R{\bar 3}$ and $C2/m$ structures by performing \textit{ab initio} calculations and strong coupling expansion. We also use symmetry considerations to limit the exchange parameters. In Sec. III, we present the transition temperatures for $R{\bar 3}$ and C$2/m$ by employing CMC simulations.
We then focus on the $R{\bar 3}$ structure and show the mechanism behind the ZZ$_1$ to ZZ$_2$ phase transition under an in-plane magnetic field in Sec. IV.
Additionally, we demonstrate that intermediate magnetic phases characterized by a large unit cell periodicity arise from interlayer interactions. In the last section, we summarize our results and discuss the implications of our results and open questions for future studies. 

\section{three-dimensional Spin model and magnetic orders}

\begin{figure} 
\includegraphics[width=1.0\linewidth,trim={0mm 00mm 0mm 00mm}]{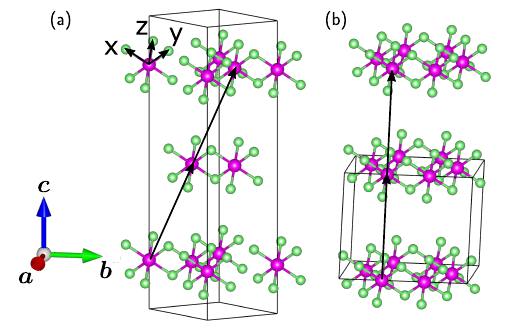}
\caption{
(a) $R\bar{3}$ and (b) $C2/m$ structure of $\alpha$-RuCl$_3$ consisting of honeycomb layers of Ru ions (purple) with edge-sharing octahedra of Cl (green). The black arrows indicate that the neighboring honeycomb layers are shifted along the $\hat{b}$ and the $\hat{a}$ axes, respectively.}
\label{Fig1}
 \end{figure}

The rhombohedral $R\bar{3}$ and monoclinic $C2/m$ structures of $\alpha$-RuCl$_3$ are shown in Fig. \ref{Fig1}.
For convenience, both structures are described in the orthorhombic coordinate system, where the $\hat{a}$ axis is perpendicular to, and the $\hat{b}$ axis is parallel to, the $z$ bonds of the honeycombs, both with the unit length equal to the bond length. 
The $\hat{c}$ axis is perpendicular to the honeycomb plane with the unit length equal to the interlayer spacing. 
The $R\bar{3}$ structure consists of honeycomb layers shifted by $(0,1,1)$ between each layer, and the layers for $C2/m$ are shifted by $(\frac{\sqrt{3}}{3},0,1)$. 

The generic spin Hamiltonian for $\alpha$-RuCl$_3$ is written as %shown in Eq. \ref{eq_hamiltonian}. 
\begin{equation}\label{eq_hamiltonian}
\mathcal{H} = H_{2D} 
+\sum_{(i,j)_n} {\bf S}_{i}^{\text{T}}\cdot \boldsymbol{\Gamma}_{c_\text{n}} \cdot {\bf S}_{j}.
\end{equation}
$\boldsymbol{\Gamma}_{c_\text{n}}$ represents the interlayer interactions for the $n$-th NN bond specific to the $R\bar{3}$ or $C2/m$ structure. 
%depicted in Fig. \ref{Fig1}.
$H_{\mathrm{2D}}$ refers to the commonly known $J-K-\Gamma-\Gamma'$ \cite{Rau2014} and the third NN Heisenberg interaction $J_3$ model \cite{Winter2016PRB}, given by
\begin{equation}
\begin{split}
H_{2D} =& \sum_{\langle ij\rangle\in\alpha\beta(\gamma)} \Big[ J{\bf S}_{i}\cdot {\bf S}_{j}+KS_{i}^{\gamma}S_{j}^{\gamma}+\Gamma(S_{i}^{\alpha}S_{j}^{\beta}+S_{i}^{\beta}S_{j}^{\alpha}) \\
&\qquad\qquad+\Gamma^{\prime}(S_{i}^{\alpha}S_{j}^{\gamma}+S_{i}^{\gamma}S_{j}^{\alpha}+S_{i}^{\beta}S_{j}^{\gamma}+S_{i}^{\gamma}S_{j}^{\beta}) \Big] \\
&+J_3 \sum_{\langle\langle\langle i,j \rangle\rangle\rangle}{\bf S}_{i}\cdot {\bf S}_{j},
\end{split}
\end{equation}
where $\langle ij\rangle$ denotes the NN magnetic sites, and $\alpha\beta(\gamma)$ denotes the $\gamma$ bond taking the $\alpha$, $\beta$ and $\gamma$ spin components in octahedral coordinate ($\alpha,\beta,\gamma\in\{\text{x,y,z}\}$). 
%$J_3$ is the intralayer third NN Heisenberg interaction. 

% The column width is: \the\columnwidth

To study the three-dimensional (3D) transition temperature $T_{N}$ and possible magnetic phase transitions under the external magnetic field at low temperatures in $\mathrm{RuCl}_{3}$, we will first investigate the forms of the interlayer interactions for both structures. Note that we will include the bond anisotropy between the $z$ and the $x,y$ bonds in the Hamiltonian $H_{\mathrm{2D}}$ later in Sec. \ref{interlayer_B}, devoted to the $C2/m$ structure case.
%since we will focus on the experimentally more interesting $R\bar{3}$ structure. The bond anisotropy are  discussed in Sec. \ref{interlayer_B}

\subsection{Interlayer spin exchange interaction for $R\bar{3}$}
\label{interlayer_A}
\begin{figure} 
\includegraphics[width=0.65\linewidth,trim={0mm 00mm 0mm 00mm}]{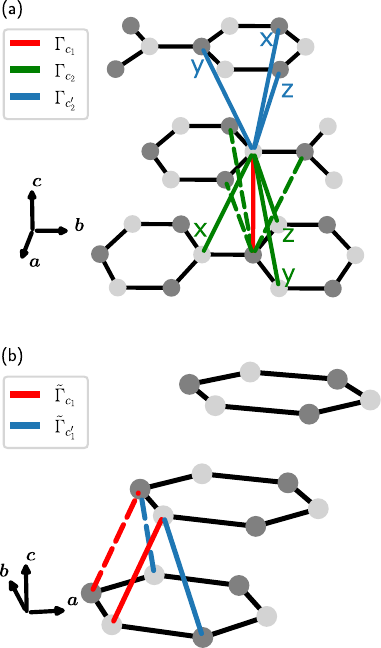}
\caption{
Interlayer interactions in the (a) $R\bar{3}$ and (b) $C2/m$ structure of $\alpha$-RuCl$_3$. Different colors represent different types of interactions. 
(a)  The $\boldsymbol{\Gamma}_{c_1}$ (red) is the NN interlayer interaction. The second NN interactions $\boldsymbol{\Gamma}_{c_2}$ (green) and $\boldsymbol{\Gamma}_{c_2^\prime}$ (blue) are related by the three-fold rotation symmetry around the $\boldsymbol{c}$ axis. The $x$, $y$, and $z$ labels indicate their bond dependence. The dashed bonds are obtained by inversion at the center of the $\Gamma_{c_1}$ (red) bond. 
(b) The first NN interlayer interactions $\boldsymbol{{\tilde \Gamma}}_{c_1}$ (red) and $\boldsymbol{{\tilde \Gamma}}_{c_1^\prime}$ (blue) in the $C2/m$ structure, where the tilde is for the $C2/m$ to differentiate from those for the $R{\bar 3}$ structure. The dashed bonds are obtained by the $\hat{a}\hat{c}$ mirror plane bisecting the $z$ bond. Since the dominant interactions are the diagonal term, they are the same as the solid bonds.}
\label{Fig2}
 \end{figure}

The geometry of the interlayer bonds is shown in Fig. \ref{Fig2} (a). Based on the lengths of the interlayer bonds between magnetic ions, there are one NN bond $\boldsymbol{\Gamma}_{c_1}$ (red line), six next-NN bonds $\boldsymbol{\Gamma}_{c_2}$ (green lines), and another three next-NN bonds $\boldsymbol{\Gamma}_{c_2^\prime}$ (blue lines) between each layer.

\subsubsection{Nearest neighbor interlayer interaction $\bf{\Gamma}_{c_1}$}

\begin{figure} 
\includegraphics[width=0.9\linewidth,trim={0mm 00mm 0mm 00mm}]{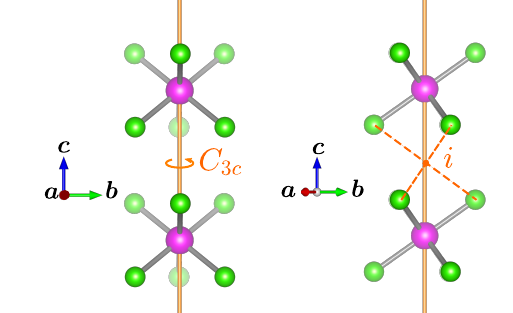}
\caption{Symmetries of the nearest neighbor (NN) interlayer bond $\boldsymbol{\Gamma}_{c_1}$ for $R{\bar 3}$. The light green Cl atoms are further behind. $C_{3c}$ is the three-fold rotational symmetry along the bond ($\boldsymbol{c}$ axis), and $i$ is the inversion symmetry at the center of the bond. These symmetries restrict the form of the $\boldsymbol{\Gamma}_{c_1}$ matrix as shown in Eq. \ref{eq_Jc1}.
}
\label{Fig3}
\end{figure}

Let us first study the spin interaction between the red bond $\boldsymbol{\Gamma}_{c_1}$ and see how it affects the 3D magnetic orders.
The three-fold rotational symmetry $C_{3c}$ along $\boldsymbol{\Gamma}_{c_1}$ and the inversion symmetry at the center of the bond, shown in Fig. \ref{Fig3}, constrain the form of the spin interactions, reflected in the hopping matrix $\boldsymbol{\mathrm{T}_{c_1}}$ in the $t_{2g}$ orbitals $(d_{yz}, d_{xz}, d_{xy})$. $\boldsymbol{\mathrm{T}_{c_1}}$ is symmetric by the inversion symmetry. $C_{3c}$ transforms the orbitals as $d_{yz} \longrightarrow d_{xz} \longrightarrow d_{xy}$, so $\boldsymbol{\mathrm{T}_{c_1}}$ has the same diagonal terms as well as the same off-diagonal terms, taking the following form:
\begin{equation}
\boldsymbol{\mathrm{T}_{c_1}} = t_2 \left(\begin{array}{ccc}
\gamma & 1 & 1\\
1 & \gamma & 1\\
1 & 1 & \gamma
\end{array}\right),
\end{equation}
where $\gamma = t_1/t_2$.
The direct hopping $t_{1}$ is much smaller than the hopping mediated through the Cl atoms $t_{2}$, so $\gamma$ can be neglected. 
Following the standard second-order perturbation calculation, the effective spin interaction for $\boldsymbol{\Gamma}_{c_{1}}$ in the octahedral coordinate  has the following form: 
\begin{equation} \label{eq_Jc1}
\displaystyle\boldsymbol{\Gamma}_{c_{1}}	=J_{c_1}\left(\begin{array}{ccc}
-1 & 1 & 1\\
1 & -1 & 1\\
1 & 1 & -1
\end{array}\right),
\end{equation}
where $J_{c_1} = t_{2}^{2}J_{H}/[(U-3J_{H})(U-J_{H})]$ with $J_H$ the Hund's coupling, and $\boldsymbol{S} = (S_x,S_y,S_z)$ defined in the $\hat{x}\hat{y}\hat{z}$ octahedral coordinate shown in Fig. \ref{Fig1} (a).
Note that the corresponding $J-K-\Gamma-\Gamma'$ model on the NN interlayer bond takes the values of $K_{c_{1}}=0$ and $J_{c_{1}}= -\Gamma_{c_{1}}= -\Gamma^\prime_{c_{1}}$.

It is more intuitive to look at $\boldsymbol{\Gamma}_{c_{1}}$ in the crystallographic $\hat{a}\hat{b}\hat{c}$ coordinate which can be obtained by rotating $\hat{x}\hat{y}\hat{z}$ to $\hat{a}\hat{b}\hat{c}$ \cite{Chaloupka2015PRB,Cen2022CP}. The above interaction becomes the XXZ model with $J_{X}=-2J_{Z}=-2J_{c_1}$, where $\boldsymbol{X}$ and $\boldsymbol{Z}$ are equivalent to $\hat{a}$ and $\hat{c}$.
This can be readily understood from $C_{3c}$ leaving the bond unchanged, meaning the bond-dependent interactions in the $\hat{a}\hat{b}\hat{c}$ coordinate must be zero. 

This form of $\boldsymbol{\Gamma}_{c_{1}}$ provides a different interpretation of the previously observed 3D magnetic orders in $\alpha$-RuCl$_3$ of the $R\bar{3}$ structure, since $J_{c_1}$ is always positive regardless of the hopping $t_2$.
The magnetic orders, named ZZ$_1$ and ZZ$_2$, are in-plane zigzag order with out-of-plane three- and six-layer periodicity, respectively \cite{balz_field_2021}.
Under a magnetic field along the $\boldsymbol{a}$ axis, the order transitioned from ZZ$_1$ to ZZ$_2$ at an intermediate field $\mytexttilde$6 T, before being suppressed at $\mytexttilde$7.2 T. Assuming small interlayer interactions, we can consider the effect of $\boldsymbol{\Gamma}_{c_{1}}$ by comparing the energies of the classical ZZ$_1$ and ZZ$_2$ orders.
$\boldsymbol{\Gamma}_{c_{1}}$ favors the zigzag chains to align along the $\boldsymbol{\Gamma}_{c_{1}}$ bond in the zero field, which corresponds to ZZ$_2$, opposite of what was assumed in the previous study \cite{balz_field_2021}. 
This is because  $J_{X}=-2J_{c_1}$ is always negative, and the in-plane component of the zigzag spin is larger than the out-of-plane component for spins about $35^{\circ}$ above the honeycomb plane \cite{Chaloupka2016,HSKim2016PRB,Sears2020NP}. Thus, we need to include the second NN interlayer interactions, $\boldsymbol{\Gamma}_{c_2}$ and $\boldsymbol{\Gamma}_{c_2^\prime}$, to explain the observed ZZ$_1$ in the zero field and the transition to ZZ$_2$ at a finite field. Below we explore the form of the second NN interlayer interactions.

\subsubsection{Next nearest neighbors ${\bf \Gamma}_{c_{2}}$ and ${\bf \Gamma}_{c_{2}^\prime}$}

The nine next-NN interlayer bonds are divided into two types. The six bonds connecting the same sublattices related by $C_{3c}$ and the inversion symmetry are labeled by $\boldsymbol{\Gamma}_{c_{2}}$ and denoted by the green solid and dotted arrows in Fig. \ref{Fig2}(a).
The remaining three bonds connecting different sublattices related by $C_{3c}$ are labeled by $\boldsymbol{\Gamma}_{c_{2}^{\prime}}$ and denoted by the blue arrows in Fig. \ref{Fig2}(a).
There is no symmetry constraint on the form of $\boldsymbol{\Gamma}_{c_{2}}$ and $\boldsymbol{\Gamma}_{c_{2}^{\prime}}$, so in general they both have nine different terms.

To simplify these interactions, we employ the \textit{ab initio} method and strong coupling expansion to determine the dominant terms (see Appendix \ref{apd_A} for details). 
Including only the dominant interactions, the forms of 
$\boldsymbol{\Gamma}_{c_{2}}$ for the $x$ bond is given by
\begin{equation} \label{eq_Jc2}
\displaystyle\boldsymbol{\Gamma}^x_{c_{2}}= 
K_{c_2} 
\left(\begin{array}{ccc}
1 & 0 & 0\\
0 & 0 & 0\\
0 & 0 & 0
\end{array}\right)
+D_{c_2}
\left(\begin{array}{ccc}
0 & 0 & 0\\
0 & 0 & 1\\
0 & -1 & 0
\end{array}\right),
\end{equation}
 where $K_{c_2}$ is an interlayer Kitaev interaction, and $D_{c_2}$ is a Dzyaloshinskii-Moriya (DM) interaction.
 Since ${\boldsymbol{\Gamma}}_{c_2^\prime}$ involves only half the number of bonds and does not contribute to the ZZ$_1$ and ZZ$_2$ transition, we ignore it in our minimal interlayer model. 
 Note that the ${\boldsymbol{\Gamma}}_{c2}$ and ${\boldsymbol{\Gamma}}_{c_2^\prime}$ interactions are bond-dependent similar to the Kitaev interaction. $\boldsymbol{\Gamma}_{c_2}^{y/z}$ for $y$- and $z$-bond can be obtained using $C_{3_c}$ rotation. 
While we can include more terms in $\boldsymbol{\Gamma}_{c_{2}}$,
we will focus on the two terms $(J_{c_1}$ in Eq. \ref{eq_Jc1} and $K_{c_2}$ in Eq. \ref{eq_Jc2}), as they are sufficient to reproduce the reported transition from ZZ$_1$ to ZZ$_2$ reported in Ref. \cite{balz_field_2021}.
As discussed in Sec. \ref{Sec_4} below and in Appendix \ref{apd_B}, the DM interaction $D_{c_2}$ does not affect the transition between ZZ$_1$ and ZZ$_2$, but it enhances the window of the intermediate phases under an in-plane magnetic field.

\subsection{Bond anisotropy and interlayer interaction in $C2/m$}
\label{interlayer_B}
The $C2/m$ structure has more extended $x$ and $y$ bonds than $R\bar{3}$ and different interlayer stacking, so the $x$ and the $y$ bonds have different interactions from the $z$ bonds in $H_{\mathrm{2D}}$. 
The $z$ bond is assumed to be the same as the $R\bar{3}$ because of the same bond length. The main effect of the longer $x$ and $y$ bonds is to reduce the direct hopping between $d_{xy}$ orbitals. Hence, from strong coupling expansion and exact diagonalization studies \cite{Winter2016PRB}, the dominant interactions are modified as $|K_{x/y}|>|K_{z}|$, $\Gamma_{x/y}<\Gamma_{z}$, $|J_{x/y}|<|J_{z}|$ and $J_{3x/3y}>J_{3z}$. For simplicity, we use a single parameter $\delta$ to quantify the anisotropic interactions of x and y bonds: $K_{x/y}=(1+2\delta)K_{z}, \Gamma_{x/y}=(1-\delta)\Gamma_{z}, J_{x/y}=(1-2\delta)J_{z}, J_{3x/3y}=(1+\delta)J_{3z}$.

The $C_2$ symmetry about the $z$ bond and the mirror plane bisecting the $z$ bond dictates that there are two types of NN interlayer bonds $\boldsymbol{{\tilde \Gamma}}_{c_{1}}$ and $\boldsymbol{{\tilde \Gamma}}_{c_{1}^\prime}$ for $C2/m$ as shown in Fig. \ref{Fig2}(b). The hopping matrix for each bond has the following form:
\begin{equation}
\boldsymbol{\mathrm{{\tilde T}}}_{c_1} =\left(\begin{array}{ccc}
t_1 & t_2 & t_4\\
t_2^\prime & t_1 & t_5\\
t_5 & t_4 & t_3
\end{array}\right), \;
\boldsymbol{\mathrm{{\tilde T}}}_{c^\prime_1} = \left(\begin{array}{ccc}
t_1 & t_2 & t_4\\
t_2 & t_1^\prime & t_5\\
t_4 & t_5 & t_3
\end{array}\right),
\end{equation}
where $t_2$, $t_2^\prime$, and $t_4$ are dominant.
Based on the dominant spin interactions from strong coupling expansion, the minimal models for the interlayer interactions have the following forms:
\begin{equation}
\displaystyle\boldsymbol{{\tilde \Gamma}}_{c_{1}} = {\tilde J}_{c_1}\left(\begin{array}{ccc}
0 & 0 & 0\\
0 & 0 & 0\\
0 & 0 & 1
\end{array}\right), \;
\boldsymbol{{\tilde \Gamma}}_{c_{1}^{\prime}} = {\tilde J}_{c_1^\prime}\left(\begin{array}{ccc}
0 & 0 & 0\\
0 & 0 & 0\\
0 & 0 & 1
\end{array}\right).
\end{equation}
Here, there is no bond dependence. There are only two NN bonds for $\boldsymbol{{\tilde \Gamma}}_{c_1}$ and $\boldsymbol{{\tilde \Gamma}}_{c_1^\prime}$.
 
\section{transition temperature in $R{\bar 3}$ and $C2/m$}
While there is a consensus that the Kitaev and $\Gamma$ interactions are dominant for $\alpha$-RuCl$_3$, the relative size of the Kitaev and $\Gamma$ interactions is still under debate  \cite{Takayama2015,Takagi2019NRP,Maksimov2020PRR,Rouso2024RoPP,matsuda2025arXiv,Maksimov2025}. 
Thus, we study two different sets of parameters suggested for the $R{\bar3}$ structure: %as shown in Table \ref{table1}:
one has $K=-2\Gamma$ \cite{HSKim2016PRB,wang2017PRB,winter2017NC,Maksimov2020PRR}, and the other has $K=-0.75\Gamma$ \cite{ran2017PRL,wang2017PRB}. 
For both cases, we set $\Gamma=8$ meV and $\Gamma^\prime=1$ meV \cite{LiuHuimei2022PRB} to satisfy the $\Gamma+2\Gamma^\prime$ constraint from the recent study on the electron spin resonance and terahertz experiments  \cite{Ponomaryov2017,Sahasrabudhe2020,Maksimov2020PRR,Maksimov2025}. The third NN Heisenberg interaction $J_3$ plays an important role in the $T_N$ as expected since $J_3$ stabilizes the zigzag order \cite{Winter2016PRB}. 
For the Heisenberg interaction $J$, it is known to be ferromagnetic \cite{matsuda2025arXiv}, with its strength adjusted to match the observed critical fields \cite{Maksimov2020PRR,Maksimov2025}. Appendices \ref{apd_A} and \ref{apd_B} show how the values of the interlayer interactions are determined. Note that the interlayer bonds have distances comparable to that of the third-nearest neighbor, with the largest interaction term approximately on the order of $O(0.1)$ meV.  We find that the two sets have similar phase diagrams, so only the results obtained using set 1 (Table 1) are shown in the main text, 
and the results using set 2 (Table \ref{table2}) are shown in Appendix \ref{apd_D}. Below, we first show the effect of interlayer interactions on the transition temperature $T_N$ of the ZZ order in the zero field. The effects on 3D magnetic transitions under an in-plane field are shown in Sec \ref{Sec_4}. The $g$ factor is 2.5 throughout this work \cite{yadav2016field,Chaloupka2016,Kubota2015}.

\begin{table}[h!]
\centering
\begin{tabular}{ |c|c|c|c|c|c|c|c|c|c|c|c| } 
 \hline
  & \multicolumn{5}{|c|}{$H_{\mathrm{2D}}$} & \multicolumn{3}{|c|}{$R{\bar 3}$} & \multicolumn{3}{|c|}{$C2/m$} \\
 \hline
 set & $J$ & $K$ & $\Gamma$ & $\Gamma^{\prime}$ & $J_3$ 
& $J_{c1}$ & $K_{c_2}$ & $D_{c2}$ 
& $\delta$ & ${\tilde J}_{c_{1}}$ & ${\tilde J}_{c_{1}^{\prime}}$ \\ 
 \hline
 1 & $-3.7$ & $-16$ & $8$ & $1$ & $1.8$ & $0.05$ & $0.2$ & $0.2$ 
 & $0.1$ & $-1.0$ & $-1.0$ \\ 

 % 2 & $-4.9$ & $-6$ & $8$ & $1$ & $1.8$ & $0.115$ & $0.2$ & $0.0$ 
 % & $0.1$ & $-1.0$ & $-1.0$ \\ 
 \hline
\end{tabular}
\caption{The set of spin exchange interactions (in units of meV) with interlayer interactions for $\alpha$-RuCl$_3$ studied in this paper.
The NN XXZ-type interaction, $J_{c1}$, second-NN Kitaev interaction $K_{c2}$, and the DM interaction $D_{c2}$ 
are the interlayer interactions for the $R{\bar 3}$ structure. $\delta$, ${\tilde J}_{c_{1}}$ and ${\tilde J}_{c_{1}^{\prime}}$ are the bond anisotropy and the interlayer interactions for the $C2/m$ structure.}
\label{table1}
\end{table}

\begin{figure} 
\includegraphics[width=1.0\linewidth,trim={0mm 00mm 0mm 00mm}]{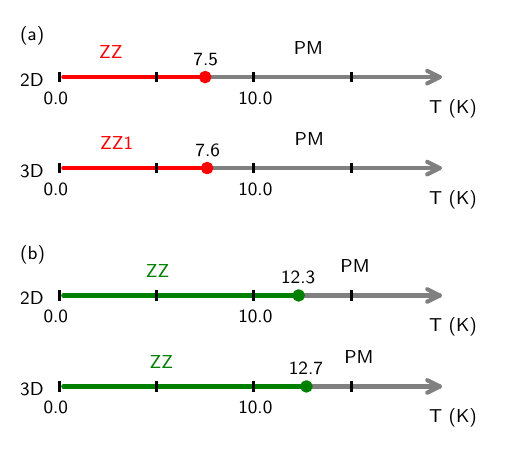}
\caption{Transition temperature $T_N$ of the zigzag order of parameter set 1 in (a) the $R{\bar 3}$ structure and (b) the $C2/m$ structure. Using parameter set 2, as shown in Appendix \ref{apd_D}, does not make a significant difference (see the main text).}
\label{Fig4_Tc}
\end{figure}

The effect of interlayer interactions on $T_N$ is shown in Fig. \ref{Fig4_Tc}.
Their impacts are rather small. The transition temperature exhibits an increase of less than $10\%$ in both $R{\bar 3}$ and $C2/m$. In contrast, the two structures show quite different transition temperatures in the 2D model due to the bond anisotropy of the $z$ bond in $C2/m$, which has a much larger impact on $T_N$. This can be understood considering the spin gap of the zigzag order, which has been estimated to be \mytexttilde 2 meV \cite{ran2017PRL}. This sizable gap due to $K$, $\Gamma$, and $\Gamma^\prime$ suppresses the fluctuation in two dimensions and stabilizes the zigzag order. The interlayer interactions have a small effect due to the large 2D magnetic anisotropy. 
On the other hand, the anisotropy of the z bond in $C2/m$ significantly enhances the 2D magnetic anisotropy. 

Although interlayer couplings have minimal effects on the transition temperature for the above parameter sets, they improve $T_N$ for models with a small spin gap. For example, when $K=-\frac{5}{4}\Gamma-\Gamma^\prime$ and $J=\frac{1}{8}\Gamma+\frac{1}{2}\Gamma^\prime$ in $R{\bar 3}$, the spin gap of the zigzag order is zero, because this model with the zigzag order can be mapped to a Heisenberg model with an antiferromagnetic order by a two-fold rotation and a four-site sublattice transformation \cite{Chaloupka2015PRB}.
Within LSWT, the gap is small when $K$ is near $-\frac{5}{4}\Gamma-\Gamma^\prime$, and $J$ has a very small impact on the gap.
Thus, we expect the finite $T_N$ to be mostly due to interlayer interactions. Indeed, our CMC simulations find that the interlayer interactions increase $T_N$ from \mytexttilde 1K to 6K, when $K$ is changed to $-11$ meV while keeping other parameters the same, as the system is closer to the hidden SU(2) symmetric point.
However, as shown above, when the model is far from such a hidden symmetric point, the interlayer couplings have minimal impacts. Despite their small effects on the transition temperature, they play an important role in field-induced phase transitions, as shown below.

\section{Three-dimensional magnetic transitions in $R{\bar 3}$} \label{Sec_4}
The importance of interlayer interactions in $\alpha$-RuCl$_3$ of the $R{\bar 3}$ structure is identified by observing the 3D magnetic transition from the ZZ$_1$ order to the ZZ$_2$ order under a magnetic field along the $\hat{a}$-axis \cite{balz_field_2021}. 
Focusing on the $R{\bar 3}$ structure, this transition provides an experimental constraint on the relative strengths of the interlayer interactions in our minimal model in Eqs. \ref{eq_Jc1} and \ref{eq_Jc2}. In Appendix \ref{apd_B}, we also show that it is also possible to explain the small critical field difference for fields along the $\hat{a}$ and $\hat{b}$ axes \cite{Maksimov2020PRR,Maksimov2025} by including more terms in the interlayer interactions. 
Here we study the minimal model in CMC simulation and show that it is able to not only produce the observed magnetic transitions, but also generate additional intermediate phases due to the small NN interlayer XXZ type $J_{c_1}$ and the second NN Kitaev interaction $K_{c_2}$ in Eq. \ref{eq_Jc2}.

\begin{figure} 
\includegraphics[width=0.99\linewidth,trim={0mm 00mm 0mm 00mm}]{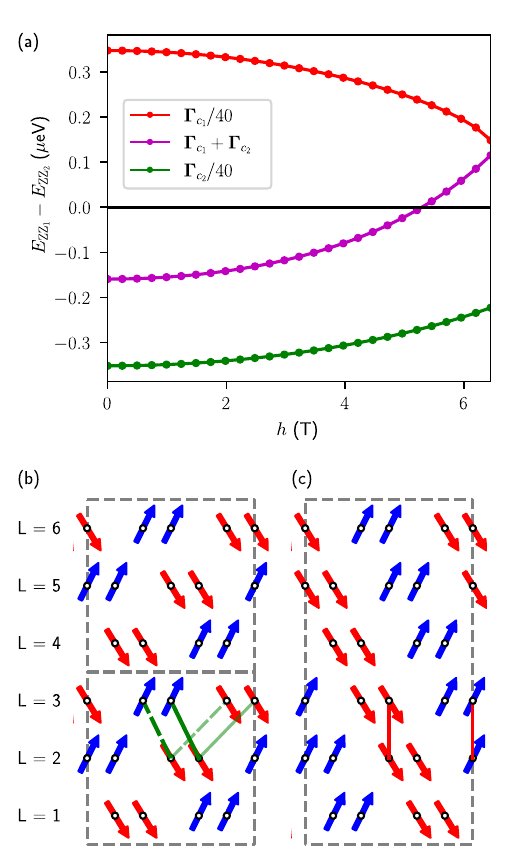}
\caption{ (a) Energy difference between 3D ZZ$_1$ and ZZ$_2$, $E_{\rm ZZ_1}$ - $E_{\rm ZZ_2}$ is shown as a function of the ${\hat a}$-axis field.
With only the NN $\Gamma_{c_{1}}$ (red), $E_{\rm ZZ_1} > E_{\rm ZZ_2}$, suggesting that ZZ$_2$ is favored, while the second-NN $\Gamma_{c_2}$ (green) and $\Gamma_{c_{2}^\prime}$ favors the ZZ$_1$ order with $E_{\rm ZZ_1} < E_{\rm ZZ_2}$. Their competition (purple) leads to the experimentally observed transition from ZZ$_1$ to ZZ$_2$. For the red and green cases, the energy difference is divided by 40 to fit the y axis window. (b), (c) Views of the ZZ$_1$ and ZZ$_2$ orders, respectively, along the zigzag chain direction. The spin directions of the two zigzag chains are drawn with blue and red arrows. The unit cells (gray rectangles) show the three- and six-layer periodicity of ZZ$_1$ and ZZ$_2$ orders. Examples of the $\Gamma_{c_2}$ bonds (green) and $\Gamma_{c_1}$ bonds (red) highlight that $\Gamma_{c_1}$ favors parallel nearest-neighbor spins, while $\Gamma_{c_2}$ favors opposite spins.
}
\label{Fig5}
\end{figure}

\subsection{ZZ$_1$ to ZZ$_2$ transition in $R\bar{3}$}
 As shown in Sec. \ref{interlayer_A} above, due to the symmetries of the $R\bar{3}$ structure, $\Gamma_{c_{1}}$ takes the form of the XXZ model with ferromagnetic interaction, Eq. \ref{eq_Jc1},
 favoring ZZ$_2$, which has parallel spins between the $\Gamma_{c_{1}}$ bond, as represented by red line in Fig. \ref{Fig5}(c). This is confirmed by the CMC calculations that show the energy difference between the ZZ$_1$ and ZZ$_2$,  $\Delta E \equiv E_{\rm ZZ_1}-E_{\rm ZZ_2}>0$ in Fig. \ref{Fig5}(a), represented by the red curve at $h=0$.

On the other hand, the second-NN $K_{c_{2}}$ favors ZZ$_1$ with $K_c>0$, as shown by the green curve with $\Delta E<0$, because the $\boldsymbol{\Gamma}_{c_{2}}$ type of interlayer interaction has more bonds connecting spins of opposite zigzag chains. This is indicated by the green lines in Fig. \ref{Fig5}(b), where two bonds (thick green line) connect opposite chains and one bond (thin green line) connects the same chain.
At low field, the effect of the second NN $K_{c_{2}}$ wins over the NN $J_{c_{1}}$ interaction, leading to ZZ$_1$, denoted by the purple curve of the energy difference obtained by the CMC simulations in Fig. \ref{Fig5}(a).
Note that the DM term $D_{c_2}$ does not affect the ZZ$_1$ to ZZ$_2$ transition, because the inversion symmetry of the $R\bar{3}$ structure is still intact in the zigzag orders (see Appendix \ref{apd_B} for details).

As the field along the $\hat{a}$-axis increases, the spins of the zigzag chains rotate toward ${\hat a}$ and get a larger common $S_a$ component, which reduces the effects of both $J_{c_{1}}$ and $K_{c_{2}}$, so $|\Delta E|$ decreases. However, the effect of $K_{c_{2}}$ diminishes faster, so $J_{c_{1}}$ wins above some critical field $h_c$ as shown in the purple curve in Fig. \ref{Fig5}(a). Therefore, when the interlayer interactions are small, the ZZ$_1$ to ZZ$_2$ transition is determined by relative strength $K_{c_{2}}/J_{c_{1}}$. A similar conclusion has been reached in Ref. \cite{balz_field_2021}, though opposite roles of ${\boldsymbol{\Gamma}}_{c_1}$ and ${\boldsymbol{\Gamma}}_{c_2}$ were assumed. Here, we provide more insight into the possible forms of the interlayer interactions based on symmetries and microscopic calculations. 

\subsection{Intermediate phases in $R\bar{3}$ under in-plane magnetic field}
\begin{figure} 
\includegraphics[width=1.0\linewidth,trim={0mm 00mm 0mm 00mm}]{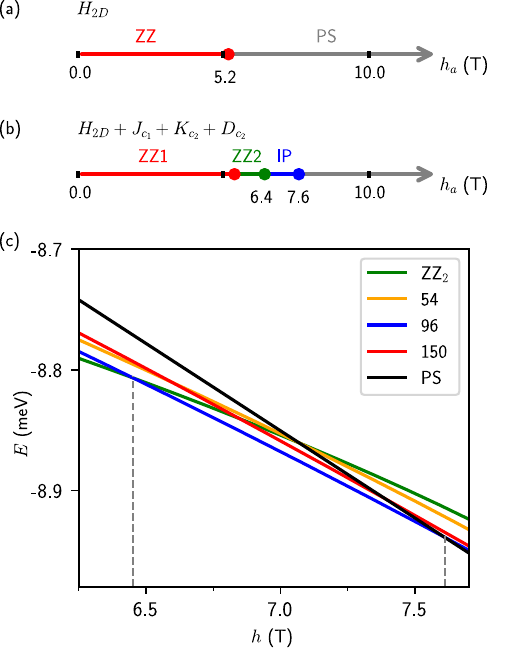}
\caption{The phase diagram under a magnetic field along the $\hat{a}$ axis for (a) 2D model and (b) 3D model including the NN and second NN interlayer interactions. The combination of the XXZ type NN ($J_{c_1}$) and the Kitaev-type second NN ($K_{c_2}$) interlayer interaction not only generates the transition between ZZ$_1$ and ZZ$_2$ but also induces an additional intermediate phase (IP) between the zigzag and the polarized phase. The second NN DM interaction ($D_{c_2}$) is not crucial for the existence of the IP, but the window of IP increases as the DM interaction strength increases (see the main text).
(c) Energies of the $N \times 1 \times N$ order versus magnetic field along the $\hat{a}$ axis near the phase transitions using the parameter set 1 in Table \ref{table1}. $N=2,3,4,5$ correspond to the ZZ$_2$ order, a 54-site order, a 96-site order, and a 150-site order, respectively. The intermediate phase with 96-site order (red) is the lowest energy state between the ZZ$_2$ and polarized states. The grey dashed lines are guides for the eyes, indicating the transitions.
}
\label{Fig6}
\end{figure}

Despite the small strength in our minimal 3D model, interlayer interactions have a significant impact on the phase diagram at the intermediate field regime. Additional IPs emerge between the zigzag phase ZZ$_2$ and the polarized phase. Using the same parameter set with the small interlayer interactions, the phase diagram under the $a$ axis field is shown in Figs. \ref{Fig6}(a) and \ref{Fig6}(b). Notably, an IP of large-unit cell order appears between the ZZ$_2$ and the polarized phase.

\begin{figure} 
\includegraphics[width=0.9\linewidth,trim={0mm 00mm 0mm 00mm}]{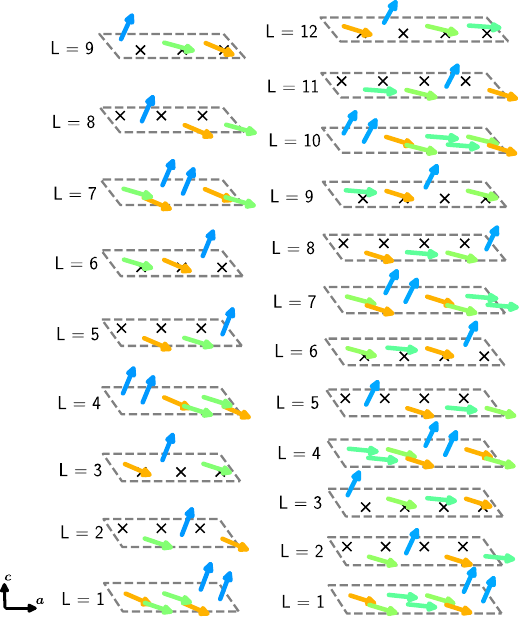}
\caption{Spin configurations of each layer of the $3\times 1 \times3$ (54-site) and the $4\times 1 \times4$ (96-site) orders. Spins with the same color have the same directions. The cross mark represents the center of a hexagon with no magnetic ion.}
\label{Apd_spin_config}
\end{figure}

To represent the IP of large unit cells, we employ
the periodicity of $N\times1\times N$, where the out-of-plane periodicity $N$ corresponds to a magnetic cell of $3N$ layers to accommodate the unit cell of the $R\bar{3}$ structure. Note that there are two sites in the in-plane primitive vectors, while three layers along the $c$ axis,  leading to $6 \times N^2$ sites in the magnetic unit cell. Figure \ref{Fig6} (c) shows the energies of the phases, with $N=2,3,4,5$ corresponding to the ZZ$_2$ order, a 54-site order, a 96-site order, and a 150-site order, respectively.
It reveals that the 96-site order ($N=4$) is the lowest energy state that emerges between the ZZ$_2$ and the polarized states as the field $h$ along the $\hat{a}$-axis increases.
 Figure \ref{Apd_spin_config} shows the spin configurations of each layer of the $3\times 1 \times3$ (9-layer) and $4\times 1 \times4$ (12-layer) orders. The $5\times 1 \times5$ (15-layer) order is not shown for brevity.

% \begin{figure} 
% \includegraphics[width=1.0\linewidth,trim={0mm 5mm 0mm 5mm}]{Fig6_IPs.pdf}
% \caption{Energies of the $N \times 1 \times N$ order versus magnetic field along the $\hat{a}$-axis near the phase transitions shown in Fig. \ref{Fig6} using the parameter set 1 in Table \ref{table1}. $N=2,3,4,5$ correspond to the ZZ$_2$ order, a 54-site order, a 96-site order, and a 150-site order, respectively. The IP with 96-site order (red) is the lowest energy state between the ZZ$_2$ and polarized states.}
% \label{Fig7}
% \end{figure}

The emergence of IPs is not unique to the specific parameter set discussed in the main text. In Appendix \ref{apd_D}, we present the phase diagram for an alternative set of exchange parameters, where two distinct large-unit-cell IPs appear between the ZZ$_2$ and polarized phases. These IPs are characterized by a 54-site and a 96-site magnetic order as shown in Fig. \ref{Apd_set2} in Appendix \ref{apd_D}. This further supports the idea that interlayer interactions play a significant role in driving the sequence of transitions observed near the partially polarized regime \cite{Lefran2023PRB, Bruin2022APL, Suetsugu2022}.

%We don't further distinguish these phases. 
%Since the additional phase is sensitive to small variation in interlayer interactions, it is natural to use interlayer interactions to explain the sample dependence of the magnetic singularities and possibly the non-monotonic behaviors in longitudinal thermal conductivity

% To further understand the origin of these additional intermediate phases, we study in-plane models. We start with the simple 2D $K-\Gamma$ model and then gradually include small intra- and interlayer interactions. The results are shown in Fig.
% We find that additional phases with periodicity $N\times 1$ (LN for $N>4$) are stabilized when the polarized state is disfavored by a positive $J$ in the 2D model. When a positive $J_3$ is added to stabilize the zigzag ground state with the desired $T_N$, a larger $J$ is required to compete with the zigzag order. 
% However, due to the positive $J$, these IPs only appear in the 2D models with very high polarizing magnetic field, which is different from the common 2D models for $\alpha$-RuCl$_3$, where a sizable negative $J$ is required to explain the moderate critical magnetic field of \mytexttilde 7T, eliminating the intermediate phases within the 2D model. 
% Interlayer interactions compete with the negative $J$ and $J_3$ interactions and stabilize the intermediate phases while maintaining a low critical field.

The IPs are 3D order arising from the cooperation between intralayer and interlayer exchange interactions through the $N\times 1\times N$ periodicity, as implied by Fig. \ref{Fig6} (c). The IPs depend on both intralayer and interlayer interactions. For example, increasing $J_3$, which enhances the ZZ$_2$ order, or increasing $|J|$, which enhances the polarized phase, can both reduce the IPs. Assuming the 96-site order in the intermediate field region, the transition temperature of the 96-site order is found to be less than 1 K. However, full calculations that include all competing phases require future investigation.

It is worth noting that IPs between the zigzag and polarized phases can also emerge in 2D models without interlayer interactions. As shown in Appendix \ref{apd_E}, certain 2D parameter sets yield IPs, such as a six-site ordered phase with a $3\times1$ periodicity, which has been previously studied \cite{Zhang2021PRB}. However, these phases are only found in 2D models with zero or positive values of $J$, resulting in an unrealistically high critical magnetic field for polarization (see Fig. \ref{Apd_fig4} in Appendix \ref{apd_E}), which differs significantly from the experimentally observed phenomenology of $\alpha$-RuCl$_3$.
To account for the critical temperature of \mytexttilde7 K as well as the moderate critical field of \mytexttilde7 T, a positive $J_3$ and a sizable negative $J$ are necessary \cite{Maksimov2020PRR}. However, such a condition tends to suppress the appearance of IPs in 2D models since the zigzag and the polarized phases are favored. Our findings suggest that interlayer interactions offer a more realistic mechanism for stabilizing IPs in $\alpha$-RuCl$_3$. 
Nonetheless, we do not exclude the possibility that alternative 2D models might also reconcile both the observed critical field strength and the presence of intermediate phases.
Therefore, our results provide a compelling and realistic route to realize intermediate phases in $\alpha$-RuCl$_3$, offering new insights into the roles of interlayer interactions and sample-dependent effects, while also leaving room for future exploration of alternative 2D scenarios and motivating studies aimed at designing ideal 2D samples.

\section{Conclusion and Discussion}
In this paper, we studied the interlayer interactions in the $R\bar{3}$ and $C2/m$ structures of $\alpha$-RuCl$_3$. We first presented that the NN interlayer spin interaction in $R\bar{3}$ is the XXZ model. Using the minimal set of interlayer interactions guided by \textit{ab initio} calculations and experimental phenomena, we showed that interlayer interactions have a small effect on the critical temperature $T_N$ of the zigzag order in the zero field. The difference in the transition temperature $T_N$ between the $R{\bar 3}$ and $C2/m$ is mainly due to the bond anisotropy of the in-plane interactions. For both structures, the interlayer couplings increase $T_N$ by less than $10\%$.

Focusing on the $R{\bar 3}$ structure, we investigated the effects of the interlayer exchange interactions on the phase transition under an in-plane magnetic field, as they may play an important role near the transitions between the zigzag and polarized states. 
% First of all, in the absence of the magnetic field, the previously observed 3D magnetic orders ZZ$_1$ and ZZ$_2$ signal the need for the interlayer interaction. 
Using the CMC simulations, we showed that the ferromagnetic first-NN interlayer interaction $(J_{c_1})$, which has the form of the XXZ model due to the $C_{3c}$ and inversion symmetry, favors the ZZ$_2$ order.  
It is the second-NN Kitaev-type interlayer interaction ($K_{c_{2}}$) that leads to the observed ZZ$_1$ order. The competition between the two interlayer interactions $J_{c_{1}}$ and $K_{c_{2}}$  gives rise to the ZZ$_1$ to ZZ$_2$ transition under a magnetic field. 
While the second-NN DM interaction $D_{c_2}$ has a similar size to $K_{c_2}$, it does not affect this transition.

The interlayer interactions not only produce the two different ZZ orders under a magnetic field but also induce additional intermediate phases before the polarized state occurs. We found that the 96-site order emerges between the ZZ$_2$ and polarized states under the in-plane $a$ axis magnetic field. 
%Given that a stacking fault represents a significant perturbation to the full 3D model, it offers a natural explanation for the sample dependence of the magnetic anomalies in $\alpha$-RuCl$_3$.

Our results suggest that full 3D models with interlayer interactions may be necessary to describe intriguing physical behaviors in $\alpha$-RuCl$_3$ such as field-induced phases. For future studies, previous analysis of experiments, such as the thermal longitudinal and Hall conductivity, may need to be revisited to include interlayer interactions.
The results presented here are obtained by CMC simulations. It is possible that the large-unit cell phases between the ZZ$_2$ and polarized states may not display a magnetic order due to quantum fluctuations in quantum models. The quantum adaptation of the 3D models presents an intriguing yet challenging problem for future research.
Increasing the interlayer interactions, magnetic orders with other larger unit cells, or even incommensurate with the lattice are possible; however, they are susceptible to the finite-size effect of the simulation, so further study is required.
Given that the intermediate phases are sensitive to variations in interlayer interactions and stacking faults, interlayer interactions offer a natural explanation for the sample dependence of the magnetic anomalies in $\alpha$-RuCl$_3$. Furthermore, it is possible that the intermediate phases arising from interlayer interactions contribute to the reported non-monotonic behavior of longitudinal thermal conductivity, an aspect that remains to be explored in future studies.

\section*{Acknowledgments}
The authors thank Y.-J. Kim and S. Kim for their useful discussions. 
This work is supported by the Natural Sciences and Engineering Research Council of Canada (NSERC) Discovery Grant No. 2022-04601. H.Y.K acknowledges support from the Canada Research Chairs Program.
Computations were performed on the Niagara supercomputer at the SciNet HPC Consortium. 
SciNet is funded by the Canada Foundation for Innovation under the auspices of Compute Canada, the Government of Ontario, Ontario Research Fund - Research Excellence, and the University of Toronto.
% This work was supported by the Natural Sciences and Engineering Research Council of Canada and the Canada Research Chairs Program. HYK thanks Aspen Center for Physics supported by National Science Foundation grant PHY-1607611, where a part of work was performed.
% %the Center for Quantum Materials at the University of Toronto. 
% This research was enabled in part by support provided by Sharcnet (\href{http://www.sharcnet.ca}{www.sharcnet.ca}) and Compute Canada (\href{http://www.computecanada.ca}{www.computecanada.ca}).
% Computations were performed on the GPC and Niagara supercomputers at the SciNet HPC Consortium. 
% SciNet is funded by: the Canada Foundation for Innovation under the auspices of Compute Canada; the Government of Ontario; Ontario Research Fund - Research Excellence; and the University of Toronto.

\appendix
\section{\textit{ab initio} calculation} \label{apd_A}
The Wannier tight-binding models for the $R\bar{3}$ and $C2/m$ structures are obtained using the maximally localized Wannier functions generated from the OPENMX codes \cite{openmx2003,openmx2004,openmx2005}. The Wannier models contain only the Ru $d$ orbitals, so the effect of the Oxygen $p$ orbitals due to strong $p$-$d$ hybridization is effectively integrated out. Below are the hopping matrices for the $\boldsymbol{\Gamma}^x_{c_{2}}$ and $\boldsymbol{\Gamma}^x_{c^\prime_{2}}$ bonds of the $R\bar{3}$ structure (in the basis $d_{yz}$, $d_{xz}$, $d_{xy}$, and in units of meV).
\begin{equation} \label{eq_hop_Jc2}
\begin{split}
\boldsymbol{\mathrm{T}}^x_{c_{2}} &= \left(\begin{array}{ccc}
-29.36 & -0.60 & -17.15\\
0.04 & 2.00 & -3.88\\
-11.27 & 9.30 & 3.50
\end{array}\right),
\\
\boldsymbol{\mathrm{T}}^y_{c_2^\prime} &= \left(\begin{array}{ccc}
-2.85 & -1.39 & -4.53\\
-1.54 & -33.96 & -12.71\\
-4.55 & -12.89 & 8.78
\end{array}\right).
\end{split}
\end{equation}

We note that the largest hopping for the $\boldsymbol{\Gamma}^x_{c_{2}}$ bond is between $d_{yz}$ and $d_{yz}$ with the strength \mytexttilde 29 meV, leading to the dominant Kitaev interaction $K_{c2}$ along the $x$ bond $\boldsymbol{\Gamma}^x_{c_{2}}$. Similarly, the other type of second-NN interlayer hopping $\boldsymbol{\Gamma}_{c^\prime_{2}}$, denoted by the blue line in Fig. \ref{Fig2}, also has a dominant Kitaev term $K_{c_2^\prime}$ due to the largest hopping ~34 meV between $d_{xz}$ and $d_{xz}$. The large hopping on the order of $O(10)$ meV, about 10\% of the NN intralayer hopping \cite{HSKim2016PRB}, gives rise to interlayer interactions of ~1\% of the large intralayer Kitaev or $\Gamma$ interactions. 

For example,  with the Coulomb interaction $U=3$ eV and the Hund's coupling $J_H=0.2U$, the resulting spin interactions (in units of meV) on the $x$ bond for the second-NN are:
\begin{equation} \label{eq_Jc2_apdx}
\begin{split}
\boldsymbol{\Gamma}^x_{c_{2}} &= \left(\begin{array}{ccc}
 0.1944&-0.0156&0.1076\\
 -0.0755&-0.0625&0.1918\\
 0.01&-0.1507&0.0066
 \end{array}\right),
\\
 \boldsymbol{\Gamma}^y_{c_{2}^\prime} &= \left(\begin{array}{ccc}
-0.047&0.0297&-0.0338\\
 0.0298&0.2005&0.0298\\
 -0.0334&0.0341&0.1131
 \end{array}\right).
\end{split}
\end{equation}

Note that the (1,1) component $\Gamma^{xx}_{c_2}$ of the ${\bf \Gamma}_{c_2}^x$ matrix, i.e., the Kitaev interaction $K_{c_2}$ is about $0.2$ meV. Similarly, the $(2,2)$ component $\Gamma^{yy}_{c_2^\prime}$ of the ${\bf \Gamma}_{c_2^\prime}^x$ matrix, i.e., another bond-dependent Kitaev interaction $K_{c_2^\prime}$ is about $0.2$ meV. However, $K_{c_2^\prime}$ is less significant because it has half the number of neighbors as $K_{c_2}$, and its effects on the zigzag orders cancel out the effects of $K_{c_2}$, as shown below in Appendix \ref{apd_B}. 
$\boldsymbol{\Gamma}_{c_{2}}$ also has a significant DM term ($D_{c_2}$). Below in Appendices \ref{apd_B} and \ref{apd_C}, we will show that the DM term has no effect on the experimentally observed zigzag orders but enhances the window of the intermediate phase. Our minimal model contains only the most important interlayer interactions: the NN $J_{c_1}$, the second-NN Kitaev interaction $K_{c_2}$, and the second-NN DM interaction $D_{c_2}$. 
In the minimal model, we set $K_{c_2}=D_{c_2}=0.2$ meV and the appropriate $K_{c_2}/|J_{c_{1}}|$ ratio for the ZZ$_1$ to ZZ$_2$ transition, ignoring ${\bf \Gamma}_{c_2^\prime}$. The intermediate phases with large unit cells also occur for other parameter choices with different finite components.

For the $C2/m$ structure, we have 
\begin{equation}
\begin{split}
\boldsymbol{\mathrm{{\tilde T}}}_{c_{1}} &= \left(\begin{array}{ccc}
2.22 & -27.87 & -12.95\\
-31.82 & 2.22 & -3.33\\
-3.33 & -12.95 & 0.13
\end{array}\right),
\\
\boldsymbol{\mathrm{{\tilde T}}}_{c_1^\prime} &= \left(\begin{array}{ccc}
4.97 & -6.26 & -6.18\\
-6.26 & -3.97 & -27.46 \\
-6.18 & -27.46 & 2.6
\end{array}\right).
\end{split}
\end{equation}

Using the strong coupling theory, we find that the interlayer exchange interactions are given by
 \begin{equation}
 \begin{split}
 \boldsymbol{{\tilde \Gamma}}_{c_{1}} &= \left(\begin{array}{ccc}
 -0.01 & 0.036 & 0.048\\
 0.038 & -0.01 & 0.05\\
 0.05 & 0.048 & -0.15
\end{array}\right),
\\
 \boldsymbol{{\tilde \Gamma}}_{c_{1}^\prime} &= \left(\begin{array}{ccc}
-0.01&0.02&0.02\\
 0.02&0.0&0.025\\
 0.02&0.025&-0.11
 \end{array}\right).
 \end{split}
\end{equation}

The interlayer interactions in the $C2/m$ structure are relatively weak. In our minimal model, we deliberately set ${\tilde\Gamma}_{c1}$ and ${\tilde\Gamma}_{c_1'}$ at 1 meV -- significantly larger than the estimated values -- to examine the impact of interlayer coupling on the transition temperature. However, even with this overestimated interlayer coupling strength, its effect remains small. Thus, we conclude that the difference in $T_N$ between the $R\bar{3}$ and $C2/m$ structures is mainly due to the anisotropy in the intralayer bonds.

\section{experimental constraints on forms of the interlayer couplings }
\label{apd_B}
%$\boldsymbol{\Gamma}_{c_{2}}$ and $\boldsymbol{\Gamma}_{c_{2}^\prime}$}

\subsection{3D magnetic orders in $R\bar{3}$}

% As the magnetic field increases, we found that ZZ$_1$ is suppressed and ZZ$_2$ occurs, as the effects of $J_{c_1}$ overcome the effects of $J_{c_2}$ and $J_{c_2}'$. 

% Thus, $\boldsymbol{J}_{c_{2}}$ and $\boldsymbol{J}_{c_{2}^{\prime}}$ are required to explain ZZ$_1$ in the zero field and the transition to ZZ$_2$ in a field. The next NN interactions need to favor ZZ$_1$ and win over $\boldsymbol{J}_{c_{1}}$ at a low field $h<h_{c}$, while they are suppressed and lose to $\boldsymbol{J}_{c_{1}}$ at a high field $h>h_{c}$.

To study what interactions in $\boldsymbol{\Gamma}_{c_{2}}$ and $\boldsymbol{\Gamma}_{c_{2}^{\prime}}$ give rise to the observed orders, we first divide them into symmetric and anti-symmetric parts (DM terms). 
% Move this to appendix: To further simplify these interactions, we first show that 
The DM terms of $\boldsymbol{\Gamma}_{c_{2}}$ and $\boldsymbol{\Gamma}_{c_{2}^{\prime}}$ can be ignored for now since they do not affect the zigzag magnetic orders for the following reasons.
When $\boldsymbol{\Gamma}_{c_{2}}$ and $\boldsymbol{\Gamma}_{c_{2}^{\prime}}$ connect the same type of zigzag chains between layers, they are merely cross product between parallel spins, so the effect is zero. 
Furthermore, when $\boldsymbol{\Gamma}_{c_{2}}$ and $\boldsymbol{\Gamma}_{c_{2}^{\prime}}$ connect different types of zigzag chains, they still have no effect, despite the spins becoming nonparallel under a magnetic field. Due to the periodicity of the zigzag order, there is always a pair of bonds with the opposite effect in the unit cell, so the net effects of the DM terms are zero.

%As argued in the main text, the anti-symmetric parts don't affect the physics of the ZZ$_1$ to ZZ$_2$ transition, so 
Let us now study each term in the symmetric part of $\boldsymbol{\Gamma}_{c_{2}}$ and $\boldsymbol{\Gamma}_{c_{2}^{\prime}}$, as shown in Eq. \ref{eq_Jc_full}, on the classical level:
\begin{equation} \label{eq_Jc_full}
\begin{split}
\boldsymbol{\Gamma}^x_{c_{2}}	&=\left(\begin{array}{ccc}
K_{c_{2}}+J_{c_{2}} & \Gamma_{c_{2}}^{xy} & \Gamma_{c_{2}}^{xz}\\
\Gamma_{c_{2}}^{xy} & K_{c_{2}}^{\prime}+J_{c_{2}} & \Gamma_{c_{2}}^{yz}\\
\Gamma_{c_{2}}^{xz} & \Gamma_{c_{2}}^{yz} & J_{c_{2}}
\end{array}\right),
\\
\boldsymbol{\Gamma}^y_{c_{2}^{\prime}}	&=\left(\begin{array}{ccc}
K_{c_{2}^{\prime}}+J_{c_{2}^\prime} & \Gamma_{c_{2}^{\prime}}^{xy} & \Gamma_{c_{2}^{\prime}}^{xz}\\
\Gamma_{c_{2}^{\prime}}^{xy} & K_{c_{2}^{\prime}}^{\prime}+J_{c_{2}^{\prime}} & \Gamma_{c_{2}^{\prime}}^{yz}\\
\Gamma_{c_{2}^{\prime}}^{xz} & \Gamma_{c_{2}^{\prime}}^{yz} & J_{c_{2}^{\prime}}
\end{array}\right).
\end{split}
\end{equation}
Since the interlayer interactions are much smaller than the intralayer interactions, we can assume that the spin directions are unaffected by the interlayer interactions, so the ZZ$_1$ and ZZ$_2$ orders can be constructed simply by stacking 2D zigzag orders. 
The ZZ$_1$ to ZZ$_2$ transition is studied by examining $\Delta E(\boldsymbol{J}_{c_1}) + \Delta E(\Gamma_{c}^{ij})$ under a magnetic field along the $\hat{a}$ axis, where $\Delta E(\Gamma_{c}^{ij}) = E_{\rm ZZ_1}(\Gamma_{c}^{ij})-E_{\rm ZZ_2}(\Gamma_{c}^{ij})$ is the energy difference with only $J_{c_1}$ or only one term $\Gamma_c^{ij}$ in $\boldsymbol{\Gamma}_{c_2}$ or $\boldsymbol{\Gamma}_{c_{2}^{\prime}}$. 
Each term satisfies $\Delta E(\Gamma_{c_{2}}^{ij})+\Delta E(2\Gamma_{c_{2}^{\prime}}^{ij})=0$ since $\boldsymbol{\Gamma}_{c_{2}}$ and $\boldsymbol{\Gamma}_{c_{2}^{\prime}}$ connect opposite zigzag chains and $\boldsymbol{\Gamma}_{c_{2}}$ has twice as many bonds. 
$K_{c_2}^\prime$ and $\Gamma_{c_{2}}^{yz}$ have the same effect as $K_{c_2}$ and $\Gamma_{c_{2}}^{xz}$, respectively, due to the $a-c$ mirror symmetry of the model without $\boldsymbol{\Gamma}_{c_{2}}$.

\begin{figure} 
\includegraphics[width=1.0\linewidth,trim={0mm 5mm 0mm 5mm}]{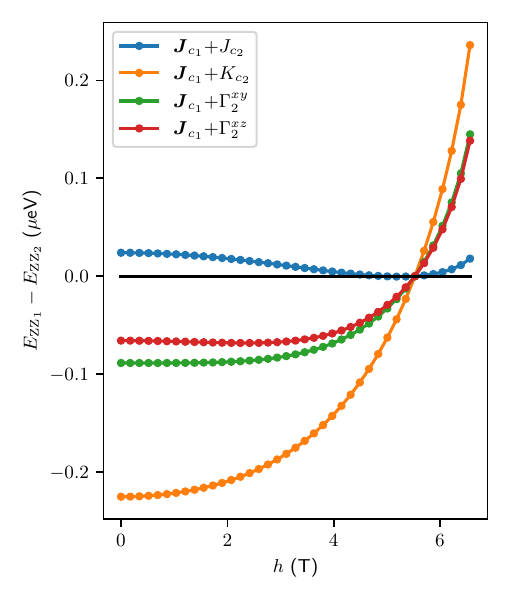}
\caption{The effect of each term in the second-NN interlayer interaction $\boldsymbol{\Gamma}_{c_2}$ on the transition from ZZ$_1$ to ZZ$_2$.}
\label{Apd_fig1}
\end{figure}

To obtain the ZZ$_1$-ZZ$_2$ transition, we must have $\Delta E(\boldsymbol{J}_{c_{1}})+\Delta E(\Gamma_{c}^{ij})<0$ at a low field and $\Delta E(\boldsymbol{J}_{c_{1}})+\Delta E(\Gamma_{c}^{ij})>0$ at a high field. The dominant Kitaev interaction $K_{c_2}$ due to the largest hopping in $\boldsymbol{\mathrm{T}}^x_{c_{2}}$ (Eq. \ref{eq_hop_Jc2}) satisfies this condition, as shown in Fig. \ref{Apd_fig1}. The relative strength ${J}_{c_{1}}$ and $\Gamma_c^{ij}$ is tuned to obtain a transition at \mytexttilde 5.5 T. It is possible to achieve the transition with a combination of other small interactions, but for simplicity we use $K_{c_{2}}$ alone to account for the transition. 
This result is not sensitive to the choice of 2D parameter sets: the ZZ$_1$-ZZ$_2$ transition can be present in various 2D parameter sets with the zigzag order, provided a suitable relative strength of $K_{c_2}/|J_{c_{1}}|$.

\subsection{Critical fields in $R\bar{3}$}
The symmetries and the observed 3D magnetic orders of the $R\bar{3}$ structure tell us the minimal form of the interlayer interactions, i.e., ${J}_{c_1}$ and $K_{c_2}$ in Eqs. \ref{eq_Jc1} and \ref{eq_Jc2}. One can include more terms in $\boldsymbol{\Gamma}_{c_2}$ to explain the critical fields in $\alpha$-RuCl$_3$.

Maksimov and Chernyshev pointed out that the difference in the critical fields for the field along the $\hat{a}$ axis $h_c^{(a)}$ and the $\hat{b}$ axis $h_c^{(b)}$ is an important constraint to the effective spin models for $\alpha$-RuCl$_3$ \cite{Maksimov2020PRR}. To explain the small difference in the critical fields, $h_c^{(a)}\approx 0.9h_c^{(b)}$, a substantial positive $\Gamma^\prime \approx \Gamma/2$ was suggested for the 2D models. 
% However, such a large $\Gamma^\prime$ contradicts the first-principle calculations for $\alpha$-RuCl$_3$ \ref{}. 
Here, we find that the small interlayer interactions play a significant role in explaining observed critical fields without invoking the above $\Gamma^\prime$ condition. 

The full 3D models are studied in CMC simulations. For the minimal model with the relative strength $K_{c_2}/|J_{c_{1}}|$ determined by the ZZ$_1$-ZZ$_2$ transition, we find that the dominant interaction $K_{c_2}>0$ also reduces the critical field difference $\Delta h_c = h_c^{(b)} - h_c^{(a)}$. 
However, a large value is required to fit the experimental critical fields. To avoid this, other interactions are included to reduce $\Delta h_c$ but do not affect the ZZ$_1$-ZZ$_2$ transition. 
For example, the following form for $\boldsymbol{\Gamma}_{c_2}$ and $\boldsymbol{\Gamma}_{c_2^\prime}$, Eq. \ref{eq_apd_Jc2}, has subdominant terms from \textit{ab initio} calculations and does not affect the ZZ$_1$-ZZ$_2$ transition since the effects of $K_{c_{2}}$ and $K_{c_{2}}^{\prime}$ mostly cancel:
\begin{equation} \label{eq_apd_Jc2}
\displaystyle\boldsymbol{\Gamma}^x_{c_{2}}=\Gamma_{c_2}\left(\begin{array}{ccc}
1 & 0 & 0\\
0 & -1 & 0\\
0 & 0 & 0
\end{array}\right),\;
\boldsymbol{\Gamma}^x_{c_{2}^{\prime}}=\Gamma_{c_2^\prime} \left(\begin{array}{ccc}
-1 & 0 & 0\\
0 & 1 & 0\\
0 & 0 & 0
\end{array}\right).    
\end{equation}
We find that with $\Gamma_{c_2}=\Gamma_{c_2^\prime}=0.4$ in addition to $J_{c_1}=0.07$ and $K_{c_2}=0.2$, $h_c^{(a)}\approx 0.9h_c^{(b)}$ is obtained.
The strength of the interaction can be further reduced if more terms are added, such as the following:
\begin{equation}
\displaystyle\boldsymbol{\Gamma}^x_{c_{2}}=\Gamma_{c_2}\left(\begin{array}{ccc}
1 & 0 & 1\\
0 & -1 & -1\\
1 & -1 & 0
\end{array}\right),\;
\boldsymbol{\Gamma}^x_{c_{2}^{\prime}}=\Gamma_{c_2^\prime} \left(\begin{array}{ccc}
-1 & 0 & -1\\
0 & 1 & 1\\
-1 & 1 & 0
\end{array}\right).    
\end{equation}
With these forms, $\Gamma_{c_2}=\Gamma_{c_2^\prime}=0.2$ is enough to get the desired $\Delta h_c$.
Our results suggest that $\alpha$-RuCl$_3$ is better described by full 3D models with interlayer interactions, since the interactions in 3D models are more consistent with first-principle calculations, suggesting a small $\Gamma^\prime$ \cite{LiuHuimei2022PRB}.

\begin{figure} 
\includegraphics[width=1.0\linewidth,trim={0mm 5mm 0mm 5mm}]{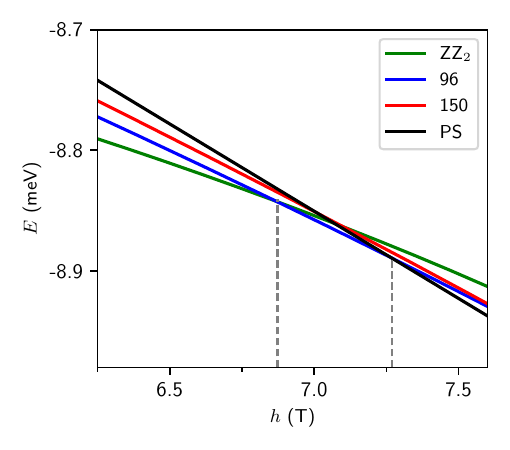}
 \caption{Intermediate phases without the DM term ($H_{2D}+J_{c_1}+K_{c_2}$). The intermediate phases still persist, though the window is smaller. The 54-site order is not shown since it is no longer a local minimum after \mytexttilde7 T. The grey
dashed lines are guides for the eyes for the transitions.
}
\label{Apd_fig2}
\end{figure}

\section{effect of DM term on intermediate phases} \label{apd_C}
As discussed above, the DM term $D_{c_2}$ in $\boldsymbol{\Gamma}_{c_{2}}$ (Eq. \ref{eq_Jc2}) does not affect the observed ZZ$_1$-ZZ$_2$ transition, so we also consider the case without $D_{c_2}$. As shown in Fig. \ref{Apd_fig2}, the window of the intermediate $N\times 1 \times N$ phases is smaller but still exists. Thus, the NN XXZ-type $J_{c_1}$ and the second-NN interlayer Kitaev interaction $K_{c_2}$ are sufficient to produce the observed 3D orders and also generate the intermediate phases.

\section{Transition temperature and magnetic fields of parameter set 2}
\label{apd_D}
\begin{figure} 
\includegraphics[width=0.9\linewidth,trim={0mm 00mm 0mm 00mm}]{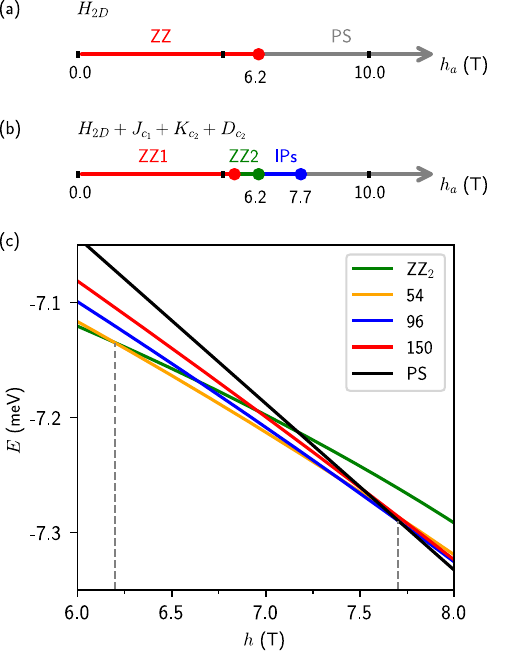}
\caption{The phase diagram under the $a$ axis magnetic field using the exchange parameter set 2 for (a) 2D model and (b) 3D model. There are two intermediate phases that emerge between the ZZ$_2$ and polarized states, as shown in (c). The lowest energy states are 54-site and then 96-site, as the field approaches the transition to the polarized state. The grey dashed lines are guides for the eyes, indicating the transitions.}
\label{Apd_set2}
\end{figure}

\begin{table}[h!]
\centering
\begin{tabular}{ |c|c|c|c|c|c|c|c|c| } 
 \hline
  & \multicolumn{5}{|c|}{$H_{\mathrm{2D}}$} & \multicolumn{3}{|c|}{$R{\bar 3}$} \\
 \hline
 set & $J$ & $K$ & $\Gamma$ & $\Gamma^{\prime}$ & $J_3$ 
& $J_{c1}$ & $K_{c_2}$ & $D_{c2}$ \\ 
 \hline
 2 & $-4.8$ & $-6$ & $8$ & $1$ & $1.8$ & $0.1$ & $0.2$ & $0.2$  \\ 

 % 2 & $-4.9$ & $-6$ & $8$ & $1$ & $1.8$ & $0.115$ & $0.2$ & $0.0$ 
 % & $0.1$ & $-1.0$ & $-1.0$ \\ 
 \hline
\end{tabular}
\caption{Parameter set 2 of the $R{\bar 3}$ structure with a smaller Kitaev interaction $|K|$. $|J|$ is increased to produce similar critical fields.}
\label{table2}
\end{table}

Since the Kitaev interaction for $\alpha$-RuCl$_3$ can take on a range of values, we also show the critical magnetic fields of parameter set 2 in Table \ref{table2} with a smaller Kitaev interaction. The transition temperature $T_N$ in the zero field is 7.3 K.
Figure \ref{Apd_set2} shows the phase diagram under the $\hat{a}$-axis field using the parameter set 2.
The 2D model exhibits a phase diagram similar to set 1 presented in the main text. For the 3D model, despite a much smaller Kitaev interaction used in this case, the presence of the IP is robust. 
The two distinct large-unit cell IPs appear between the ZZ$_2$ and polarized phases in the 3D model. These IPs are characterized by the 54-site and
the 96-site magnetic orders as shown in Fig. \ref{Apd_set2}(c). This further supports the importance of the interlayer interactions in determining the sequence of transitions observed near the partially polarized regime. 

\section{Intermediate phases in two-dimensional models}
\label{apd_E}
\begin{figure} 
\includegraphics[width=1.0\linewidth,trim={0mm 00mm 0mm 00mm}]{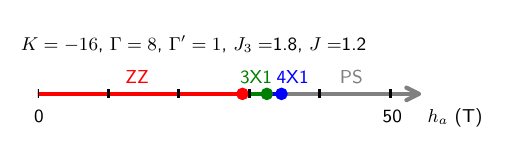}
\caption{The phase diagram of a 2D model under a magnetic field along the $\hat{a}$ axis. The intermediate phases arise from the competition between $J$, $J_3$, and $\Gamma$ and appear with a positive $J$ but at a large field. All the parameters are in units of meV.
}
\label{Apd_fig4}
\end{figure}

The 2D $JK\Gamma \Gamma^\prime J_3$ model can also generate intermediate phases after the zigzag order is suppressed under a magnetic field along the $\hat{a}$-axis, as shown in Fig. \ref{Apd_fig4}. 
Intermediate phases with large unit cells have been previously studied, such as the $3\times1$ (six-site) phase \cite{Zhang2021PRB}. They arise from the competition between $J$, $J_3$, and $\Gamma$. A positive $J_3$, which stabilizes the zigzag order in a low field, can overcome the intermediate phases. However, if $J$ is also increased, then the intermediate phases can be stabilized again in a higher field. 
Thus, when a large $J_3$ is needed to reach the transition temperature of the zigzag phase $T_N$~7 K, the required magnetic field for intermediate phases is very high (~30 T for $\Gamma$~10 meV) due to the positive $J$ and $J3$.

% With only $|K|/\Gamma=2$, the ground state has $3\times1$ periodicity (6-site order in Ref. \cite{Zhang2020}) before the polarized state. With the addition of a positive $J=0.1 |K|$, states with larger unit cells appear. The $3\times1$ and the large-unit cell states persist when a small $J_3$ is added to stabilize the zigzag order at low field. If $J_3$ is too large, we only have the zigzag order before the polarized state. 
% If we add a positive $\Gamma^\prime$ and use a larger $J_3$ to stabilize the zigzag order, the $4 \times 1$ order appears following the $3\times1$ order.

In contrast, $\alpha$-RuCl$_3$ has a moderate critical field of ~7 T, which usually requires a sizable negative $J$ and makes it difficult to generate IPs if only 2D models are used.
%The interlayer interactions compete with the negative $J$ and positive $J_3$ interactions, stabilizing the intermediate phases with a low critical field. 
However, as we stated in the main text, we cannot rule out alternative
2D models that reconcile both the observed critical field strength and the presence of intermediate phases in pure 2D systems.

\bibliography{references}

\end{document}